\documentclass[preprint,superscriptaddress,prl,longbibliography]{revtex4-1}
\usepackage{graphicx}
\usepackage{amsmath,amssymb}
\usepackage{mathrsfs}
\usepackage[normalem]{ulem}
\usepackage{color}
\usepackage{soul}
\usepackage{hyperref}
\usepackage{textcomp}

\usepackage[resetlabels, labeled]{multibib}

\usepackage[T1]{fontenc}
\usepackage[utf8x]{inputenc}
\usepackage{bm}
\usepackage[left]{lineno}

\newcommand{\be}{\begin{equation}}
\newcommand{\ee}{\end{equation}}
\newcommand{\bea}{\begin{eqnarray}}
\newcommand{\eea}{\end{eqnarray}}

\newcommand{\ignore}[1]{}

\newcommand{\UFMGDCC}{Computer Science Department, Universidade Federal de Minas Gerais, Belo Horizonte, MG 31270-901, Brazil}
\newcommand{\UFMGCM}{Microscopy Center, Universidade Federal de Minas Gerais, Belo Horizonte, MG 31270-901, Brazil}
\newcommand{\UFMGEE}{Department of Electronic Engineering, School of Engineering, Universidade Federal de Minas Gerais, Belo Horizonte, MG 31270-901, Brazil}
\newcommand{\UFMGDF}{Physics Department, Universidade Federal de Minas Gerais, Belo Horizonte, MG 31270-901, Brazil.}

\newcommand{\UFMGPIn}{Technology Innovation Graduate Program, Universidade Federal de Minas Gerais, Belo Horizonte, MG 31270-901, Brazil.}

\newcommand{\UFMGPGEE}{Electrical Engineering Graduate Program, Universidade Federal de Minas Gerais, Belo Horizonte, MG 31270-901, Brasil.}

\newcommand{\UFBA}{Instituto de Física, Universidade Federal da Bahia,
Campus Universitário de Ondina, Salvador - BA, 40170-115 Brazil.}
\newcommand{\NIMS}{National Institute for Materials Science (NIMS), 1-2-1 Sengen, Tsukuba-city, Ibaraki 305-0047, Japan.}

\renewcommand{\phi}{\varphi}
\renewcommand{\epsilon}{\varepsilon}

\begin{document}

\title{The limits of Near Field Immersion Microwave Microscopy evaluated by imaging bilayer graphene Moir\'{e} patterns}

\author{Douglas A. A. Ohlberg}
\affiliation{\UFMGCM}

\author{Diego Tami}
\affiliation{\UFMGPGEE}

\author{Andreij C. Gadelha}
\affiliation{\UFMGDF}

\author{Eliel G. S. Neto}
\affiliation{\UFBA}

\author{Fabiano C. Santana}
\affiliation{\UFMGDF}

\author{Daniel Miranda}
\affiliation{\UFMGDF}

\author{Wellington Avelino}
\affiliation{\UFMGPGEE}

\author{Kenji Watanabe}
\affiliation{\NIMS}

\author{Takashi Taniguchi}
\affiliation{\NIMS}

\author{Leonardo C. Campos}
\affiliation{\UFMGDF}

\author{Jhonattan C. Ramirez}
\affiliation{\UFMGPGEE}
\affiliation{\UFMGEE}
\author{C\'{a}ssio Gon\c{c}alves do Rego}
\affiliation{\UFMGPGEE}
\affiliation{\UFMGEE}
\author{Ado Jorio}
\affiliation{\UFMGPGEE}
\affiliation{\UFMGPIn}
\affiliation{\UFMGDF}
\author{Gilberto Medeiros-Ribeiro*}
\affiliation{\UFMGPGEE}
\affiliation{\UFMGDCC}

\maketitle

\textbf{Molecular and atomic imaging required the development of electron and  scanning probe microscopies to surpass the physical limits dictated by diffraction~\cite{Abbe1881}. Nano-infrared experiments~\cite{Chen2012} and pico-cavity tip-enhanced Raman spectroscopy imaging later demonstrated that radiation in the visible range~\cite{Lee2019} can surpass this limit by using scanning probe tips to access the near-field regime~\cite{Synge1928}. Here we show that ultimate resolution can be obtained by using scanning microwave imaging microscopy to reveal structures with feature sizes down to 1~nm using a radiation of 0.1~m in wavelength. As a test material we use twisted bilayer graphene, which is not only a very important recent topic due to the discovery of correlated electron effects such as superconductivity~\cite{Zondiner2020}, but also because it provides a sample where we can systematically tune a superstructure Moiré pattern´s modulation from below one up to tens of nanometers. By analyzing the tip-sample distance dynamics, we demonstrate that this ultimate 10$^8$ probe-to-pattern resolution can be achieved by using liquid immersion microscopy concepts and exquisite force control exerted on nanoscale water menisci.  
}

\pagebreak

Liquid immersion microscopy has its roots in observations made by Hooke in 1667~\cite{hooke1667} that images would improve in clearness and brightness upon spreading fluids onto the surface and gently moving it upwards towards the lens until touching. Additionally, the adhesion of liquid to the lens was so robust and firm that it bore the investigated object being moved about the field of view. This vivid description of meniscus formation and usage was subsequently expanded in 1813 with Brewster's concept of the oil immersion lens~\cite{Bradbury1967}. Later, in 1855, Amici improved upon several construction aspects, concerned primarily with diminishing the loss of light in high power microscopes by opting for water as the immersion liquid~\cite{Bradbury1967}. Ensuing developments that further addressed the issues of light loss and improvement of the magnification power of lenses consolidated the recognition of immersion lens microscopy as a well established technique. 

Albeit remarkable, all these developments are diffraction limited, defined by the Abb\'{e}'s resolution limit of $d={\lambda}/2n{\sin{\theta}}$, with $n\sin{\theta}$, as the numerical aperture. The proposal of scanning aperture imaging by Synge \cite{Synge1928} for near field imaging was put into practice in 1972 by Ash \cite{ASH1972}, improving magnification beyond the Abb\'{e} limit with a figure of merit of $\lambda/d$ of 60. Operation in the near field regime has since enabled a tremendous advance in microscopy, deserving a detailed and fair review that falls outside the scope of the present letter, as it would encompass implementations at many different wavelengths, construction details and application fields. In the microwave regimen there are interesting opportunities to be explored, as the field is at the crossroads of optics and electronics. 

Scanning microwave impedance microscope (sMIM) is one of the latest additions to the family of scanning probe microscopes. Commercially available \cite{primenano} tools can now be used to retrofit existing equipment, and exciting results in multiple applications have been published \cite{Amster2017, Seabron2019, Lee2020} describing exquisite spatial detail and vector analysis of the microwave reflected signal at each pixel. A 3~GHz microwave signal is coupled to an Atomic Force Microscope (AFM) probe tip that works as a waveguide and performs as an apertureless near-field microscope~\cite{Amster2017}. A key differentiating aspect of sMIM is that, unlike Scanning Tunneling Microscopy, its ability to image nano-scale modulations in the electronic and dielectric properties of complex structures is not restricted to conductive samples, but possible with insulating dielectrics as well. A point worth emphasizing, is that the capacitance signal conveys dielectric, geometric, and quantum information. Previously, Seabron \cite{Seabron2019} employed sMIM to assess quantum capacitance of carbon nanotubes. Capacitance spectroscopy is a technique historically employed to map the electronic and quantum properties of quantum dots~\cite{Ashoori1992, Alegre2006} and two-dimensional quantum systems~\cite{Yu2013}. In these systems, a dielectric layer is mandatory for a proper adjustment of the chemical potential between probe electrode (gate, tip) and system of interest (quantum structure). Seabron~\cite{Seabron2019} noted that to improve the spectroscopic resolution of sMIM, a high pemittivity (high-k) capping layer would be essential to better couple the tip to the sample, and that adventitious water found on surfaces forming a meniscus could be an option. With $\epsilon_r=\epsilon/ \epsilon_0\approx 80$ and a refractive index of 9 at 3~GHz frequencies, its surface ubiquity demands that the effect of water must be included in any modeling of sMIM experiments at ambient conditions.

\begin{figure*}[!hbtp]
\centering
\centerline{\includegraphics[width=183mm]{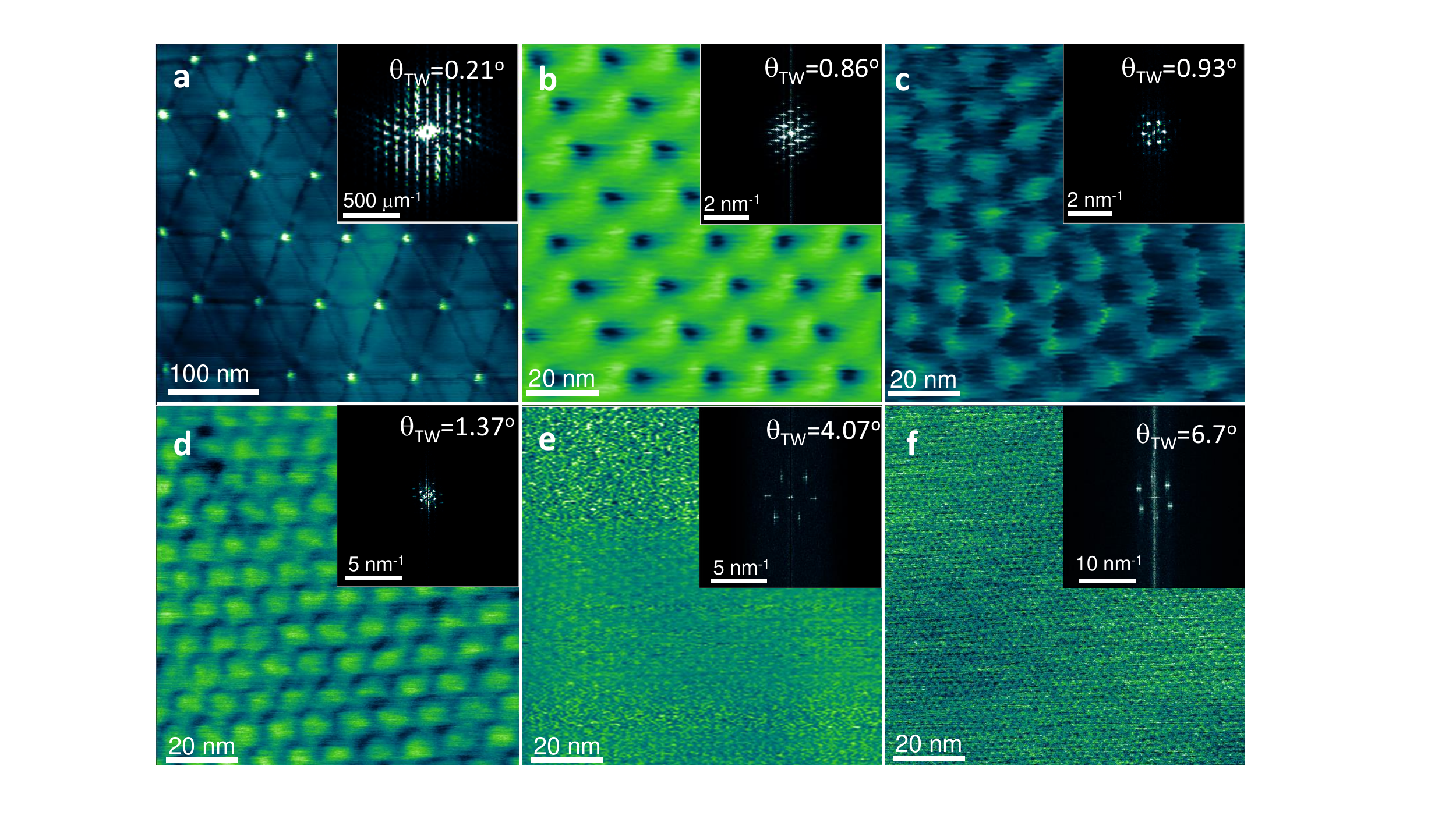}}
\caption{\textbf{sMIM scans}. \textbf{a}. 400~nm x 400~nm conductance  image of TBG:hBN:glass, with contrast arising from the juxtaposing of two graphene layers with $\theta_{TM}$= 0.21$^\circ$, and the strain soliton domain wall arising from surface reconstruction clearly resolved~\cite{McGilly2020}. The false color is keyed to the intensity of reflected signal, white being the highest, i.e., higher conductance. The observed pattern is consistent with a reconstructed structure. The inset shows the corresponding FT. \textbf{b-f}. 100~nm x 100~nm scans exhibting a wide range of angles and corresponding Moir\'{e} patterns. The systems examined are TBG:hBN:glass (\textbf{a}, \textbf{b}, \textbf{e}, \textbf{f}), TBG:SiO$_{2}$:Si~\textbf{c},  and TBG:glass~\textbf{d}.}
\label{fig1}
\end{figure*} 

Here we report sMIM results with unprecedented spatial resolution performed on Twisted Bilayer Graphene (TBG) systems of varying twist angles. TBG systems offer an exciting opportunity to create two-dimensional superlattices of varying periodicity in a conceptually simple strategy of adjusting the twist angle $\theta_{TM}$ between the two graphene layers. The search for systems producing two-dimensional periodic potential modulation in appropriate dimensions has seen a variety of implementations over the years, for example ranging from antidot lattices~\cite{Ensslin1990} and top-gate modulation~\cite{Schlsser1996} in 2D electron gases in III-V heterostructures in the 1990's, to more recent and exciting TBG embodiments \cite{Dean2013}. The possibility to explore the potential modulation parameters in a more detailed fashion offered by the van der Wall heterostructures such as graphene presented surprising opportunities near the magical angle of $\approx 1.1^{\circ}$ that goes beyond metal-insulator transition and Wigner crystallization~\cite{Schlsser1996}, allowing the observation of additional electron-correlation physics such as superconductivity~\cite{Zondiner2020}.  Tools that can expeditiously analyze and provide answers on the electronic structure, preferably at ambient conditions, are currently being pursued~\cite{McGilly2020, Lee2020}. 

Figure \ref{fig1} shows a series of sMIM scans over a set of twisted bilayer graphene (TBG) systems, with twist angles $\theta_{TM}$ of \textbf{a.} 0.21$^\circ$, \textbf{b.} 0.86$^\circ$, \textbf{c.} 0.93$^\circ$, \textbf{d.} 1.37$^\circ$, \textbf{e.} 4.07$^\circ$ and \textbf{f.} 6.7$^\circ$. These samples were  characterized by Raman spectroscopy, Tip Enhanced Raman Spectroscopy microscopy~\cite{gadelha2020lattice} and UHV-STM (see Materials and Methods Section for all the experimental details), to independently verify the bilayer locations and confirm the observed Moir\'{e} superlattices that arise in TBG systems. TBG systems are atomically flat, which conveniently eliminates surface topography contributions to the reflected microwave signal, leaving  the underlying electronic and dielectric structure components in the admittance intact. Figure~\ref{fig1} demonstrates the ability of the sMIM to observe the solitonic structures that arise in the atomically reconstructed TBGs prepared with twist angles smaller than 1.1$^\circ$~\cite{McGilly2020} (figure~\ref{fig1}~\textbf{a}) and many interesting features as we move through the relaxation transition (figures~\ref{fig1}~\textbf{b.} through \textbf{f.}), thus establishing  sMIM as an important tool to non-invasively examine these systems in ambient conditions. The false color scale is keyed to the intensity of the real part of the reflected microwave signal, i.e., the conductance component. Inspection of figure~\ref{figS2} shows how inextricably connected are the complex reflected microwave to the complex admittance of the sample. All the data shown here are non-filtered, and the only image processing performed was background removal and color range adjustment. In the upper right corner of each image, a Fourier Transform (FT) of the data is displayed, showing diffraction spots corresponding to the periodic modulation of the electronic properties due to the Moir\'{e} two-dimensional superlattice. The sequence spans a wide range of periods, culminating in a 6.7$^\circ$ twist angle and a period $1/f$ of 2.1~nm for the sample in \textbf{f}. 

The Nyquist frequency $f_c$, defined by the highest frequency that can be inferred from a signal requires a spatial resolution of at least $2f_c$~\cite{proakis}. Thus, our resolution is better than 1.05~nm ($1/2f$), despite the fact that the tip radius is 50 times bigger. Considering the Abb\'{e}'s limit, our figure of merit is 10$^8$. This unprecedented resolution requires a deeper investigation.

\begin{figure*}[!hbtp]
\centering
\centerline{\includegraphics[width=183mm]{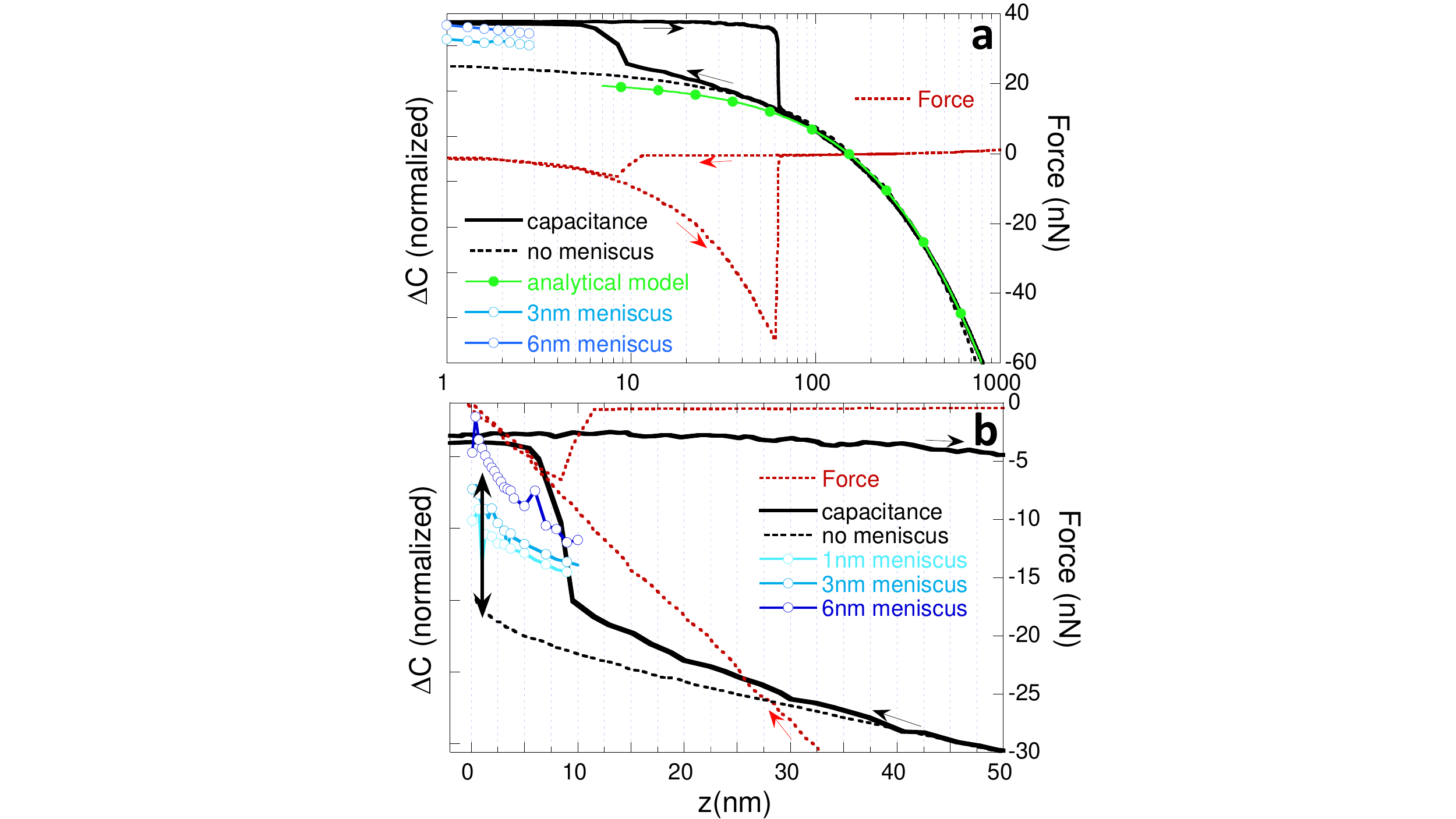}}
\caption{\textbf{a}. The reference of $z=0$ is defined on tip contact, which happens after the tip pull in by the meniscus, and subsequent motion $z$-piezo until the deflection reaches zero force. Negative $z$ corresponds to pushing the tip into the repulsive mode. A typical force-curve approach of a WTi metallic tip towards the TBG layer. In red the cantilever deflection and corresponding force. The black solid line corresponds to the measured capacitance signal. The green line with solid circles correspond to a fit of the data to equation \ref{eq1}, with the tip radius fixed at 50nm. The black, dashed line represents the FEM model assuming a 1nm water layer on both surfaces. The blue empty circles correspond to the FEM model of the system assuming a water meniscus of 3nm and 6nm radii, respectively, with the thickness as the independent variable z. \textbf{b}. Close up of the same data, in a linear plot in the vicinity of meniscus formation and snap off. The arrows indicate the tip direction, with the corresponding colors for capacitance (black) and force (red).}
\label{fig2}
\end{figure*} 

The reflected microwave signal is a complex function that depends on substrate admittance (sample conductivity and permittivity, see figure~\ref{figS2} in supplemental material section), with the imaginary part related to the system capacitance. Examining the tip-surface approach curves tracing cantilever deflection and capacitance signal allows us to assess tip-surface mechanical and electrical coupling. The observed capacitance and force behavior during tip approach and retraction with respect to a TBG:hexagonal Boron Nitride (hBN):Glass stack are shown in black and red dotted lines in figures~\ref{fig2}~\textbf{a},~\textbf{b}. The directions of approach and retraction on the capacitance data are indicated by arrows. At about $\sim$10~nm sample-substrate distance, the tip experiences capillarity attraction ($\sim$7~nN force) and the capacitance jumps. At the onset of the tip deflection due to the attractive force of the meniscus, the x-axis no longer represent tip-surface distance but rather z-piezo displacement because of the cantilever elastic deformation towards the surface.

The data are fit to an analytical model for the capacitance between a tip and a surface~\cite{Hudlet1998}, described by equation~\ref{eq1}:

\begin{equation}
C_{\mathrm{meas}}=C_{\mathrm{stray}}+2\pi \varepsilon _{0}R\ln[1+R(1-\sin \theta _{0})/z]
\label{eq1}
\end{equation}  

with $C_{\mathrm{stray}}$ as the stray capacitance, $\varepsilon _{0}$ the vacuum permittivity, $R$ as the tip radius, $\theta _{0}$ the aperture angle, chosen to be about $10^\circ$~\cite{Hudlet1998},  and $z$ the tip height. The data and fit (thin green line with solid circles) are plotted as  $\Delta C=C(z)-C({1\,\mu\mathrm{m}})$, and multiplied by a normalizing constant. The tip radius $R$ was kept fixed at 50~nm, its nominal value. The agreement between data and analytical model captures the capacitance dependence on $z$ from 1~$\mu$m to about 50~nm from the surface.  

To accurately describe the capacitance of a tip in close proximity to the surface, we performed Finite Element Method (FEM)~\cite{volakis1998} modeling using  the COMSOL\texttrademark ~Multiphysics simulation tool taking into account the contribution of both adventitious water and meniscus formation to the measured capacitance (see details in suppl. materials, figure~\ref{figS1}).  Water is ubiquitous and frequently considered an unwanted nuisance that complicates nano-scale phenomena, although it has been used as a resource (see, for example~\cite{Cheng2011}). Its role in the theory of capillary forces has also been exhaustively studied ever since  Hooke’s work with water menisci, and his observations can today be precisely assessed with any commercially available or home made microscopes to the degree that understanding of capillarity has advanced to the nanoscale.  AFM embodies one of the most convenient tools to probe capillarity at the nanoscale level. Experiments covering meniscus dynamics formation~\cite{Szoszkiewicz2005} and meniscus stiffness~\cite{Carpentier2015} illustrate the level of control and understanding that has been achieved of the tip-water meniscus-surface system.  This body of knowledge can be used to engineer menisci at the nanoscale, since precise control of force and tip velocity~\cite{Szoszkiewicz2005},  and environmental conditions can be done routinely nowadays. 

The capacitance dependence on $z$ derived from the FEM model is shown as a black dashed line for a system with adventitious water of 1~nm on the tip and surface, and empty circles in different shades of blue for the additional contribution of water meniscus of 2-6~nm radii~\cite{Szoszkiewicz2005} and varying thickness. The model captures the capacitance dependence on $z$ for most of the experimental range, except at the onset of meniscus formation. This excess capacitance can be accounted for by the meniscus nucleation process. In the literature, the proposed values for meniscus thickness are of the order of 0.2~nm in close proximity~\cite{Bartok2017} to 1.65~nm at snap-off~\cite{Pakarinen2005}. Calculated values of meniscus of constant radii and decreasing thickness ($z$ axis) at the 0.2-1.65~nm in blue circles in figure~\ref{fig2}\textbf{b} show a reasonable agreement with the measured capacitance jump (indicated by the double arrow). Within the simplifying assumptions for the proposed geometry, amount of water on the surfaces, dynamics of menisci formation, and capacitive forces~\cite{Hudlet1998} that may pull the tip closer, the capacitance behavior derived from the simulated model captures the essence of the tip approach and meniscus formation and closely matches the experimental data and analytical model. A more detailed evolution of tip retraction and capacitance decrease of the tractioned meniscus~\cite{Carpentier2015} is shown in supplemental figures~\ref{figS3}~\textbf{a},\textbf{c}.

The dynamic aspect of meniscus formation \cite{Szoszkiewicz2005}, which for tip approach is of the order of a few ms, impacts not only the force-curve approach curves but scanning generally speaking. In fact, for the scan rates employed (1~$\mu$m/s for figure~\ref{fig1}.c) the estimated meniscus radius is 2~nm \cite{Szoszkiewicz2005}. As an additional test to corroborate the presence of a meniscus, we performed experiments in the so called NAP mode (see supplemental material), which basically is a set of two consecutive line scans, one in non-contact and the second at a pre-defined lift height. We were able to continue imaging at lift heights of up to 50~nm piezo displacement (see figure \ref{figS3} in the supplemental material section), with a minor capacitance drop ($\sim$0.01$\Delta C$) and sustained meniscus presence. We never observed any image contrast in AC driven intermittent contact (or tapping) mode imaging in our experiments, which could be due to a poor or nonexistent meniscus formation (tip oscillation frequency of 70~kHz, and therefore a period of 1.4~$\mu$s which could be too short \cite{Szoszkiewicz2005}). Operation in repulsive mode led to tip wear and sample damage, and not necessarily better imaging conditions. 

\begin{figure*}[!hbtp]
\centering
\centerline{\includegraphics[width=183mm]{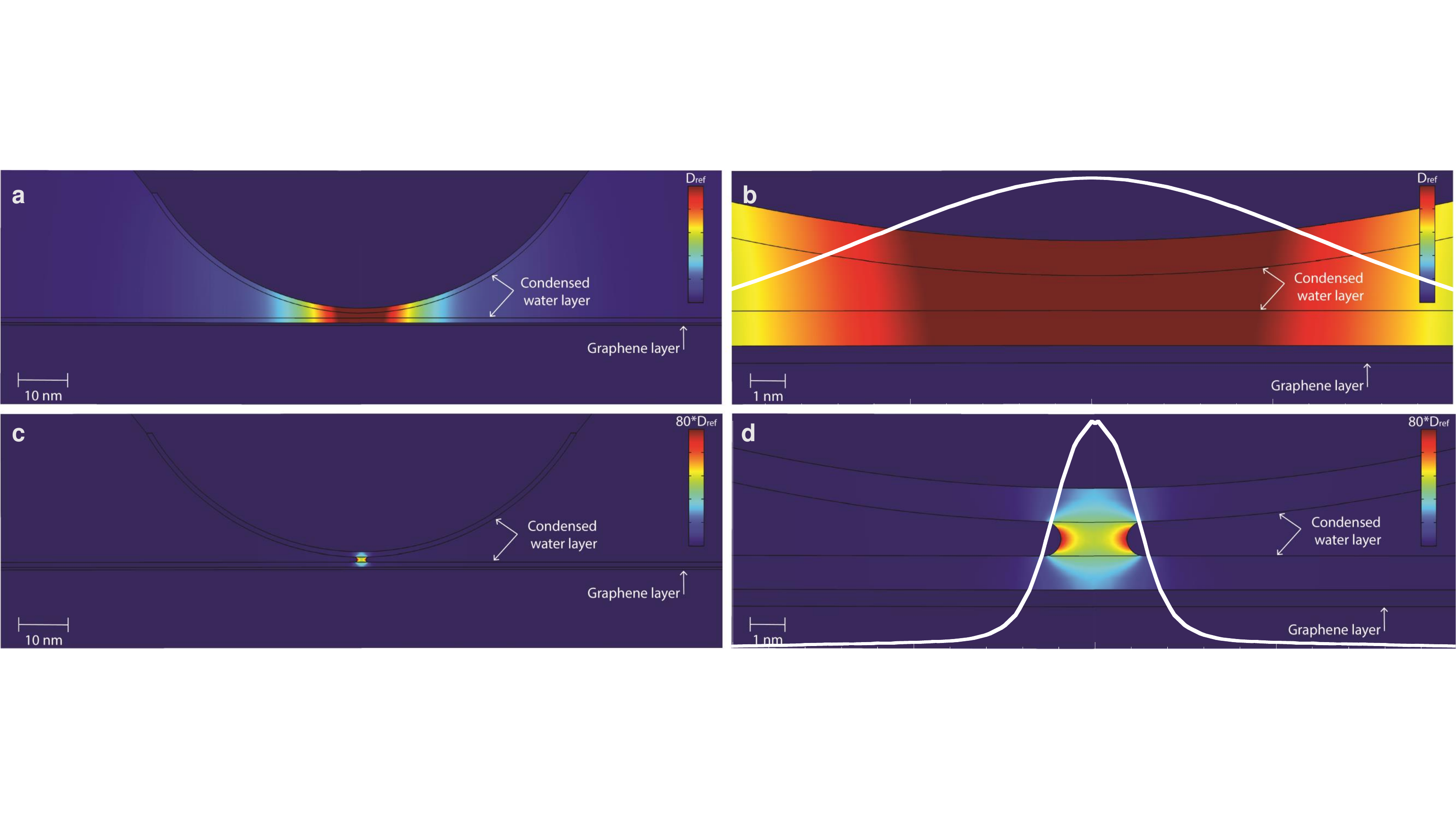}}
\caption{\textbf{Electric displacement field distribution in the tip/sample simulated structure. $\textbf{D}_{Ref} =2.5\times10^{-3}$}C/m$^2$ ~\textbf{a.} Tip/sample simulated system for Graphene:hBN:Glass substrate at 1~nm $z$ distance without meniscus, and a 1nm layer of water on both surfaces. \textbf{b.} Close-up on the structure detailed in \textbf{a}. \textbf{c.} Simulation for the same structure in \textbf{a}, with the additional implementation of a water meniscus. The color range has been expanded by $\textbf{D}_{full scale}=80~\textbf{D}_{Ref} $in order to permit the visualization of the increased density in $\textbf{D}$. \textbf{d.} Close-up the structure detailed in \textbf{c}, also with the expanded scale. Line profiles of the normalized Electric Displacement Field $\textbf{D}_{full scale}=1$ are superposed onto the image and demonstrates the concentration effect of the water meniscus.}
\label{fig3}
\end{figure*} 

In Figure~\ref{fig3} we examine the field distribution in the vicinity of and inside the water meniscus at the onset of meniscus formation (assumed to happen at 1~nm). The FEM calculated distribution of $\textbf{D}\equiv \varepsilon _{0}\textbf{E}+ \textbf{P}$ over the entire system and its evolution during tip approach can be seen in the videos \textbf{SV1} in supplemental material. The simulation results shown in figure~\ref{fig3} exhibit the configurations of tip-substrate without meniscus figures~\ref{fig3}~\textbf{a},\textbf{b}  and with meniscus figures~\ref{fig3}~\textbf{c},\textbf{d}, on top of a TBG layer. The majority of the displacement field is localized at the apex of the tip upon meniscus formation, but noteworthy is the fact that it concentrates at the water:TBG interface, as seen in the $\textbf{D}$ profiles (white lines) in figures~\ref{fig3}~\textbf{b},\textbf{d}. The effect of the meniscus demonstrates the ability to concentrate fields in very small regions, corroborating the modeled and experimentally observed capacitance jumps. From a microwave perspective, the meniscus is an iris that upon force control and operation in the attractive mode allows for impedance matching and field concentration. Thus, menisci can be used to augment near-field resolving power. 
   
Finally, rigorous methods for evaluation of resolution rely on well defined Point Spread Functions (PSF) from which one can deduce the Transfer Function of the optical system. That would allow us to verify the resolution beyond the Nyquist limit that represents an upper bound. The challenge of identifying a point-like defect to extract the ultimate resolution along with additional correlation analysis is discussed in the supplemental material section.

Near field and immersion optics at microwave frequencies create opportunities worth exploring. One convenient aspect of sMIM is the lack of externally coupled optical apparatus, enabling conectorized tools for easy deployment. Further possibilities can be envisioned for near field immersion microscopy. The water layer requirement may for instance allow examination of biological samples, using single layer talc or hBN sheets as cover-slips for adequate microwave transmission. The ability to implement spatially resolved capacitance spectroscopy by means of DC biasing schemes is an exciting prospect, becoming an invaluable tool for van der Walls heterostructures and band-gap engineering. An often explored resource of scanning probe microscopy is nanolithography. Yet, for the majority of the tools employed in nanolithography, the embodiements are normally implemented in an open loop fashion, allowing only post-mortem inspection. With the reflected microwave signal, one can close the loop and monitor the complex impedance of the region of interest, while performing the patterning~\cite{NFEdelivery2019}. An immediate implementation in the already vast field of dip-pen nanolithography~\cite{Liu2020} would envision tracking both the real and imaginary parts of the reflected microwave signal to enable real time tracking of minute quantities of dispensed materials, each with its impedance signature.

\subsection*{References}
\bibliographystyle{unsrt}
\bibliography{superhires}

\subsection*{Materials and Methods}
\textbf{Scanning microwave impedance microscopy:} The AFM used to support the SMIM acquisition was a  MFP-3D-SA manufactured by Asylum Research. The shielded co-axial AFM probes as well as the electronics unit (model Scanwave Pro) used to transmit and measure the microwave signal were manufactured by PrimeNano Inc. The experiment schematics are shown in figure~\ref{figS0} of the supplemental material section. All SMIM and AFM data were collected under ambient conditions.
Scanning tunneling microscopy (STM):  STM data were collected with a UHV VT STM/AFM model manufactured by Omicron GmbH. The tips used were etched tungsten probes. The STM was calibrated with a standard graphite lattice observed on an HOPG surface and a Si (111) reconstructed 7 X 7 surface lattice. All STM data were collected at room temperature at a pressure of 1 X 10$^{-10}$ Torr. The Moiré patterns observed by STM were obtained from TBG samples initially supported by oxide-coated Si coupons for sMIM inspection. The bilayers were then, transferred to conducting, gold-coated mica coupons for subsequent STM analysis. 

\textbf{Tip-enhanced Raman spectroscopy (TERS):} The nano-Raman system used for the TERS analysis was a hybrid configuration combining micro-Raman optics with an atomic force microscope that was assembled in-house. The system employs an inverted optical microscope that by means of an oil immersion objective (1.4 numerical aperture) tightly focuses a radially polarized HeNe excitation laser source (633 nm) onto the AFM/TERS probe and local surface being scanned by the probe. The AFM/TERS probes are especially designed to resonate with the excitation laser producing highly localized field enhancements which are then back-scattered by the analyte sample into the spectrometer and charge-couple device (CCD) that collects the Raman signal~\cite{gadelha2020lattice}.  All TERS experiments in this work were conducted under ambient conditions. 

\textbf{Preparation of TBG:} The twisted bilayer samples analyzed were prepared using a novel technique we have developed that is a variation of conventional, dry transfer, tear-and-stack methods~\cite{gadelha2020lattice}. Unlike other procedures ,which either completely encapsulate graphene bilayers within a top and bottom layer of h-BN flakes or a bottom layer of h-BN flake and a top layer of polymer that often requires removal in subsequent solvent soaks and sample bakes, our dry transfer procedure fabricates simple, extremely clean, and unencapsulated TBGs free of polymers that can introduce undesired contaminants. The procedure relies on a special stamp design consisting of a truncated, polymer pyramid fabricated on a handle substrate that is not only capable of performing tear-and-stack operations on the initial graphene, but also allows subsequent detachment of the bilayer onto a variety of support substrates. (See supplemental) The substrate supports included simple glass coverslips with and without h-BN coating layers that were used for tip-enhanced Raman spectroscopy (TERS) analysis, mica coupons coated with atomically-flat gold for STM analysis, and silicon coupons coated with an 275nm oxide.  After transfer to a respective substrate, a WITec Alpha 300 SAR confocal Raman Microscope was used for Raman spectroscopy and spatial mapping. These measurements were typically performed using a 633 nm laser, power of 5 mW, and spot size of 600 nm. 

\subsection*{Acknowledgments}
The authors acknowledge finacial support from CNPq, FINEP, FAPEMIG, INCT Nanomateriais de Carbono, and CAPES.

\subsection*{Competing financial interests}
The authors declare no competing financial interests
\subsection*{Author contributions}
\textbf{Sample preparation:} Andreij C. Gadelha, Daniel Miranda, Fabiano C. Santana, Eliel G. S. Neto, Leonardo C. Campos; K. Watababe and T. Taniguchi provide hBN crystals; \textbf{Micro-Raman and TERS measurements:} Andreij C. Gadelha, Eliel G. S. Neto; \textbf{SPM measurements:} Douglas A. A. Ohlberg, Gilberto Medeiros-Ribeiro; \textbf{Finite Element Computation:} Diego Tami, Jhonattan C. Ramirez, Cássio Gonçalves do Rego; \textbf{Microwave technical support:} Wellington Avelino, Gilberto Medeiros-Ribeiro; \textbf{Data Analysis:} Gilberto Medeiros-Ribeiro, Douglas A. A.Ohlberg; \textbf{Project idealization and guidance:} Ado Jorio, Jhonattan C. Ramirez, Cássio Gonçalves do Rego and Gilberto M. Ribeiro; \textbf{Paper writing:} Douglas A. A. Ohlberg and Gilberto Medeiros-Ribeiro. Some authors contributed with parts of the text and figures, and they all read and agreed on the final version of the manuscript.

\pagebreak
\widetext
\begin{center}
\textbf{\large Supplemental Materials}
\end{center}
\setcounter{equation}{0}
\setcounter{figure}{0}
\setcounter{table}{0}
\setcounter{page}{1}
\makeatletter
\renewcommand{\theequation}{S\arabic{equation}}
\renewcommand{\thefigure}{S\arabic{figure}}
\renewcommand{\bibnumfmt}[1]{[S#1]}
\renewcommand{\citenumfont}[1]{S#1}

\section{Experimental setup}
\begin{figure}[htbp]
  \includegraphics[width=\linewidth]{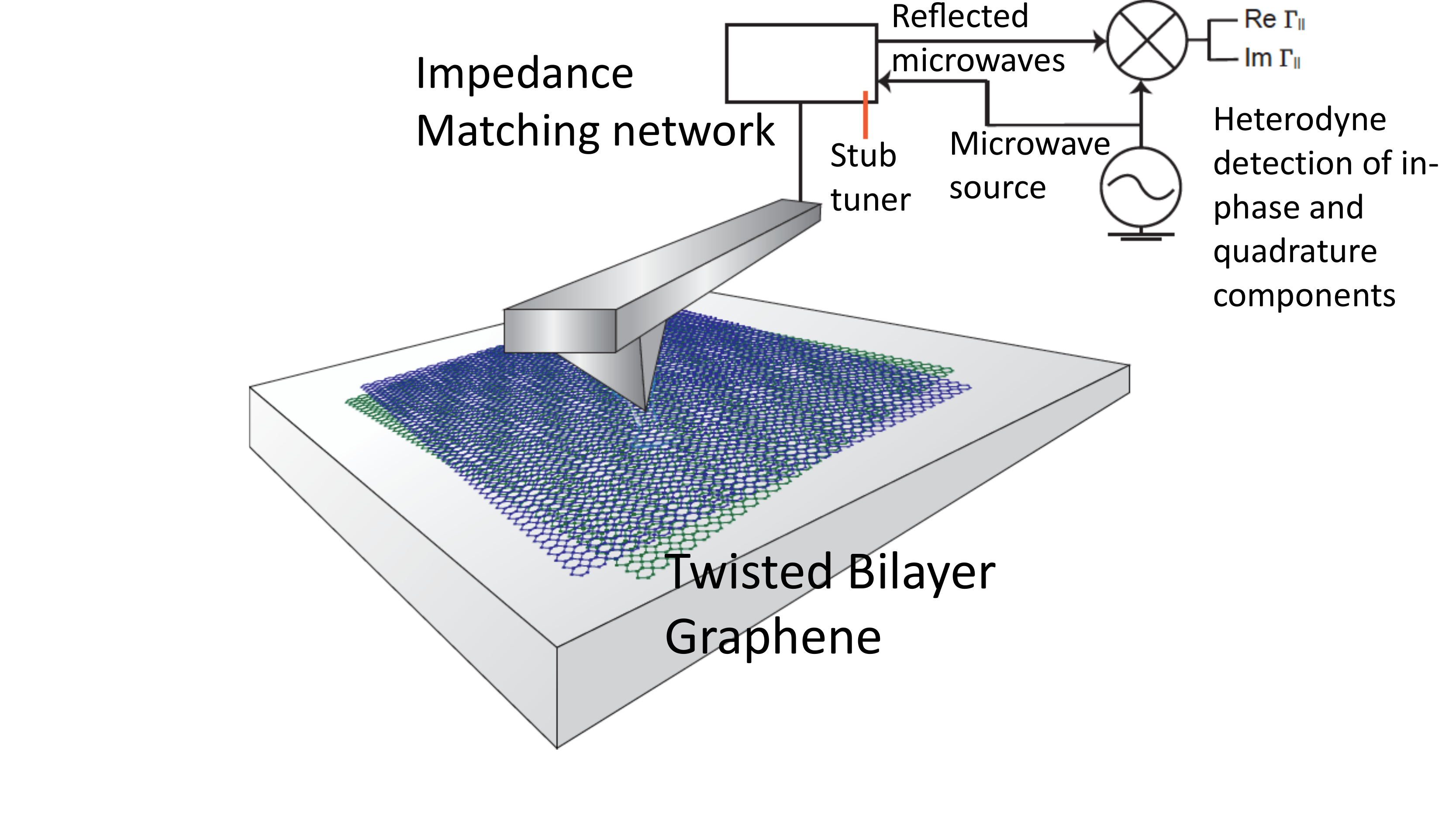}
  \caption{Experimental set up showing the microwave circuitry for sample probing.}
  \label{figS0}
\end{figure}
\section{Finite Element Method calculations}
Aiming to fully understand the effects observed in the measurements obtained by our scanning Microwave Impedance Microscopy (sMIM) system, some numerical simulations were performed implementing the Finite Element Method (FEM)~\cite{volakis1998s} in the COMSOL\texttrademark Multiphysics simulation tool, in order to calculate the tip-sample admittance.

\begin{figure}[htbp]
  \includegraphics[width=\linewidth]{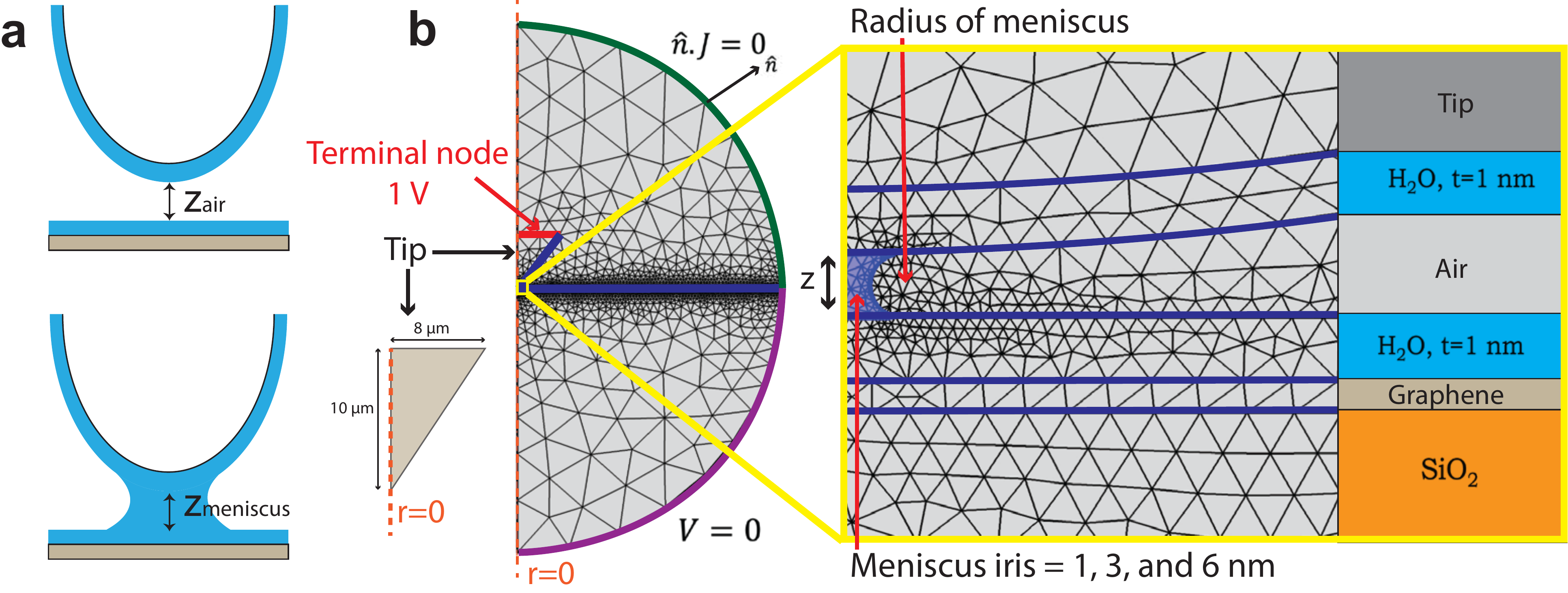}
  \caption{Simulation structure. \textbf{a}. Tip and surface/sample with and without meniscus. \textbf{b}. Scheme of highlighting boundary conditions at electrical isolation and ground potential (zero potential). Radius of the tip, 50~nm. Simulated twisted double layer graphene on SiO$_{2}$:Si, assuming a 0.5~nm thick TBG with $\sigma = 10^{6}$ S/m and a 275nm thick SiO$_{2}$ with $\epsilon_r = 4$.}
  \label{figS1}
\end{figure}

The sMIM system was simulated in an axisymetrical two-dimensional environment, i.e., taking the axial symmetry boundaries (at r = 0) into account, and consequently add an axial symmetry node to the component, which becomes valid just on the axial symmetry boundaries, figure~1, in order to reduce the computational cost of the simulation performed~\cite{koshiba1992s}. The implemented tip has an inverted pyramid shape, with $8~\mu$m at the top and $10~\mu$m in height. The structural characteristics of the implemented simulation system, can be seen in figure.~\ref{figS1}.

As presented in figure~\ref{figS1}, the gap between the tip and the sample is just a few nanometers, providing the ideal conditions for the formation of a nanometric-order bridge between the tip of the simulated sMIM and the surface, due to the capillary condensation of water in our system, as described in the main text.

In a real measurement system, the admittance varies as function of the distance $z$. As the tip approaches the surface, the condensed water particles form the meniscus, depending on tip-surface.The simulations carried out contemplate two different scenarios, with and without the presence of the water meniscus, using a mesh with ~13,000 total elements, of which ~1,500 correspond to the elements in the contour conditions.

For systems with high frequency, the imaginary component of the admittance, the susceptance contribution, becomes many orders of magnitude larger than the real component, the conductive contribution. 

\begin{equation}
C=\frac{Im(Y)}{\omega},
\end{equation}

where $Y$ is the admittance, $C$ is the capacitance, and $\omega = 2\pi f$ is the angular frequency at $f=3$~GHz.

\begin{figure}[htbp]
 \includegraphics[width=\linewidth]{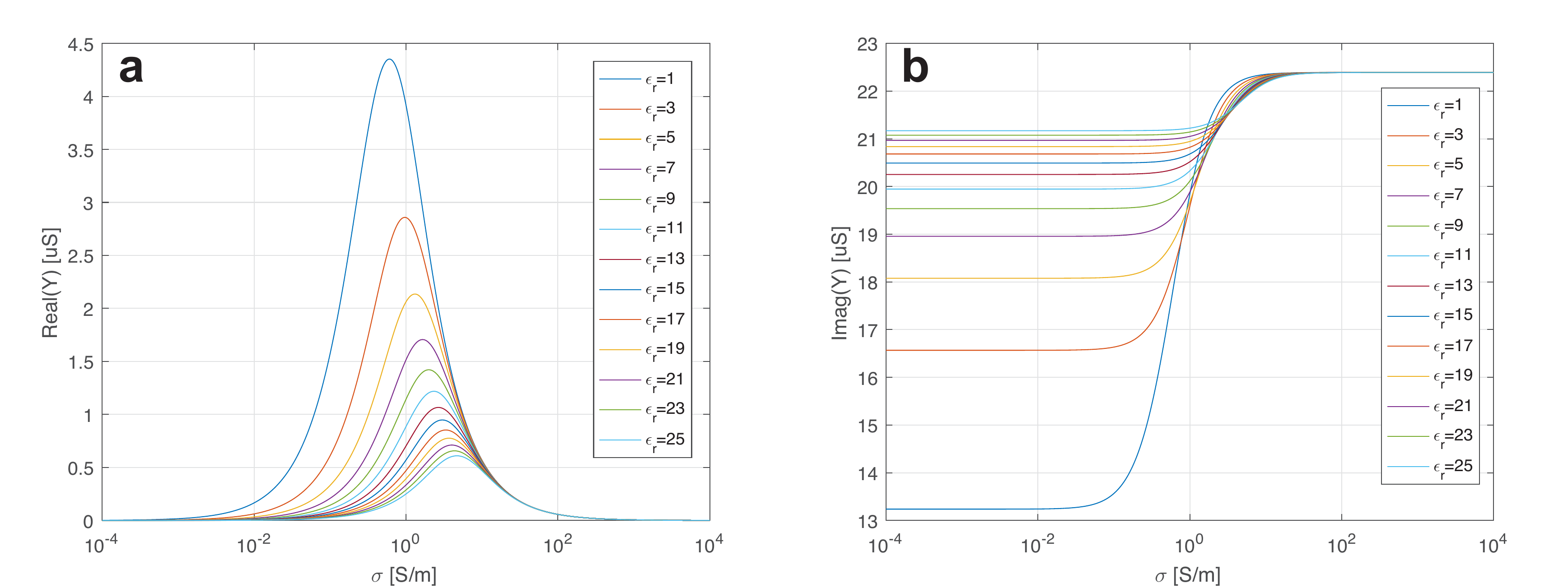}
  \caption{\textbf{a} Real and \textbf{b} Imaginary part of the admittance as a function of the conductivity and permittivity}
  \label{figS2}
\end{figure}

Finally, the real and the imaginary part of the admittance were calculated as a function of the permittivity and the conductivity of the sample, the results corresponding to 1~nm meniscus on two single layer graphene over SiO$_{2}$ can be seen in figure~\ref{figS2}.

\pagebreak
\section{Independent confirmation by Scanning Tunneling Microscopy}

\begin{figure}[htbp]
  \includegraphics[width=\linewidth]{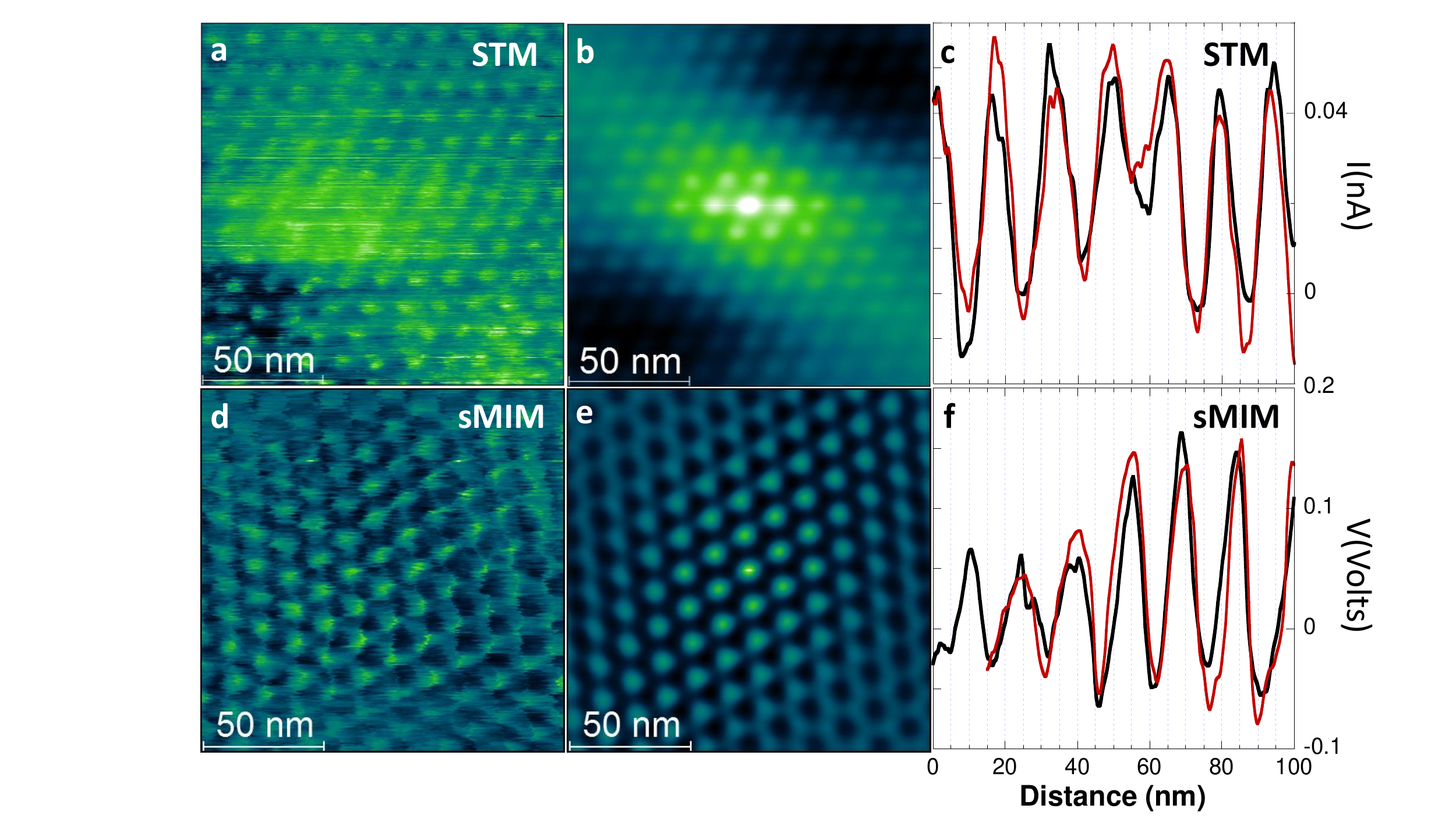}
  \caption{\textbf{Comparison between STM and sMIM data.} \textbf{a.} 150~nm x 150~nm STM current image, \textbf{b.} its corresponding autocorrelation function and \textbf{c.} line profiles along the symmetry axis, \textbf{d.} 150~nm x 150~nm sMIM conductance image, \textbf{e.} its corresponding autocorrelation function, and \textbf{f.} line profiles along the symmetry axis.}
  \label{figS3}
\end{figure}

In order to confirm the observation of the Moir\'{e} superlattices by sMIM, we perform Ultra-High Vacuum Scanning Tunneling  Microscopy (UHV-STM). The samples were transferred from the glass substrate to a Au:Mica substrate and brought into the UHV system. The precise flake location was determined by a macro lens system that allowed a 10~$\mu$m resolution. We were able to get images from regions of similar twist angles ($\theta_{TM}$=0.93$^\circ$ for the sMIM experiments and $\theta_{TM}$=1.10$^\circ$ for the STM experiments). Figure~\ref{figS3} shows the image data, autocorrelation functions and line profiles for the STM (top row) and sMIM (bottom row).

\pagebreak
\section{Discussion on PSF}

\begin{figure*}[!hbtp]
\centering
\centerline{\includegraphics[width=183mm]{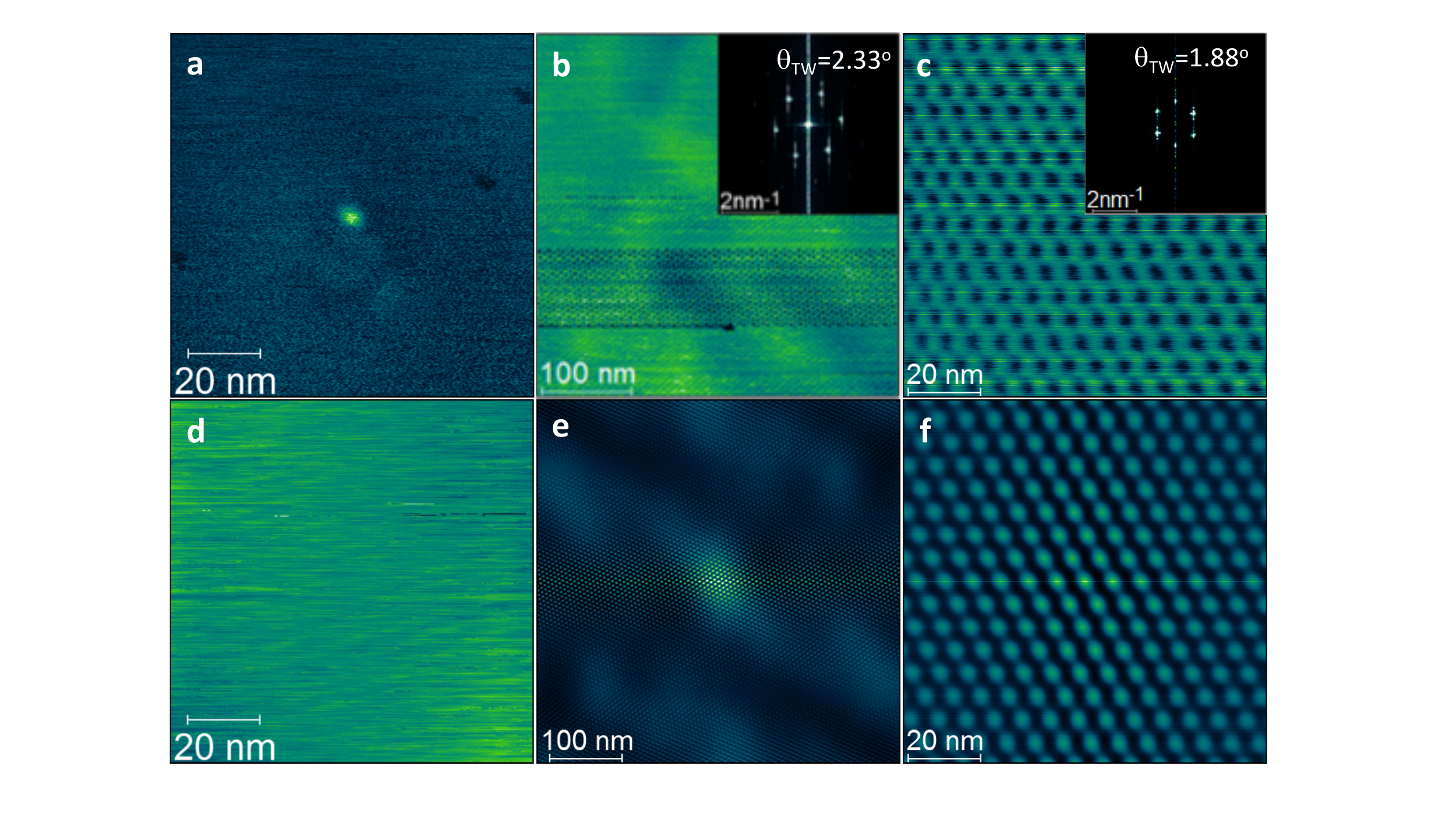}}
\caption{\textbf{a}. 100~nm x 100~nm sMIM scan with an isolated defect on a otherwise flat piece of graphene. \textbf{b}. 400~nm sMIM scan on a TBG with $\theta_{TM}=2.33^{\circ}$. \textbf{c}. 100~nm sMIM scan on a TBG with $\theta_{TM}=1.88^{\circ}$. \textbf{d}. 100~nm x 100~nm topographic image corresponding to \textbf{a}, with a 300~pm full scale color scale. \textbf{e} Autocorrelation Function (ACF) taken on image \textbf{b}, showing a long range order, and \textbf{f} ACF taken on image \textbf{c}, showing a high degree of order.}
\label{figS4}
\end{figure*} 

Figure \ref{figS4}\textbf{a} shows a 100~nm x 100~nm sMIM scan with an isolated defect on a otherwise flat piece of graphene (figure \ref{figS4}\textbf{d} as the topography channel, 300~pm color scale range). This isolated object has a full width-half maximum of about 8~nm, but still bigger than an ideal PSF, since we were able to infer a resolution better than 1.05~nm. Yet, this isolated structure permits us to rule out tip artifacts that would otherwise create higher periodicity. Figure~\ref{figS4}\textbf{b} shows a 400~nm sMIM scan on a TBG with $\theta_{TM}=2.33^{\circ}$ and Figure~\ref{figS4}\textbf{e} its corresponding autocorrelation function (ACF). Inspection of the autocorrelation function reveals minute details such as long range order and the existence of texture, shear stress and other defects, information that is essential for modeling the electronic properties of these systems. Finally, figure\ref{figS4}\textbf{c} and \textbf{f}  show a 100~nm sMIM scan on a TBG with $\theta_{TM}=1.88^{\circ}$ and its corresponding ACF. Moir\'{e} patterns had been used before in the geometrical optics context by deflectometry for the determination of the Modulation transfer Function of optical systems\cite{Glatt1985s}. For this particular system, despite the fact that the FT unequivocally shows the periodicity of 1/2.1~nm$^{-1}$, autocorrelation functions shown in figures \ref{figS4}\textbf{e} and \textbf{f} that contain essentially the same information as the FT permit nevertheless a better assessment of the imaged surface, and also how uniform and long range order present in the surface, ruling out tip artifacts. 

\pagebreak
\section{NAP experiments}

The NAP experiments have a key goal to further substantiate the presence and use of the water meniscus as an imaging device. The term “nap pass” is used by Asylum Research that refers to an expression used in  aviation “nap-of-the-earth flight,” that describes flight made at low altitude which closely tracks the contours of the underlying terrain~\cite{Proksch2016s}. The experiment is performed in two steps: 1) one pass in contact or non-contact mode in order to extract the topography; and 2) a second pass at a predetermined height $\Delta~z$ above the surface using the topography information acquired in the previous scan in order to maintain the tip-substrate distance constant. This protocol is executed in a line-by-line fashion, in order to avoid sample drift. The convenience of a flat sample is instrumental to allow a unequivocal assessment of the capacitance and conductance signal. For this particular sample, the majority of the contrast was measured in the conductance channel, indicating a sample conductivity above 10~S/m, as seen in figure~\ref{figS2}. Yet, the capacitance contains the meniscus information, and comparing its value in contact and lift mode permits us to assess some degree of meniscus deformation, rule out the possibility of a metallic fragment as the sole component responsible for the resolution, and evaluate any potential contrast change associated with meniscus deformation due to the cantilever lift.

\begin{figure}[htbp]
  \includegraphics[width=\linewidth]{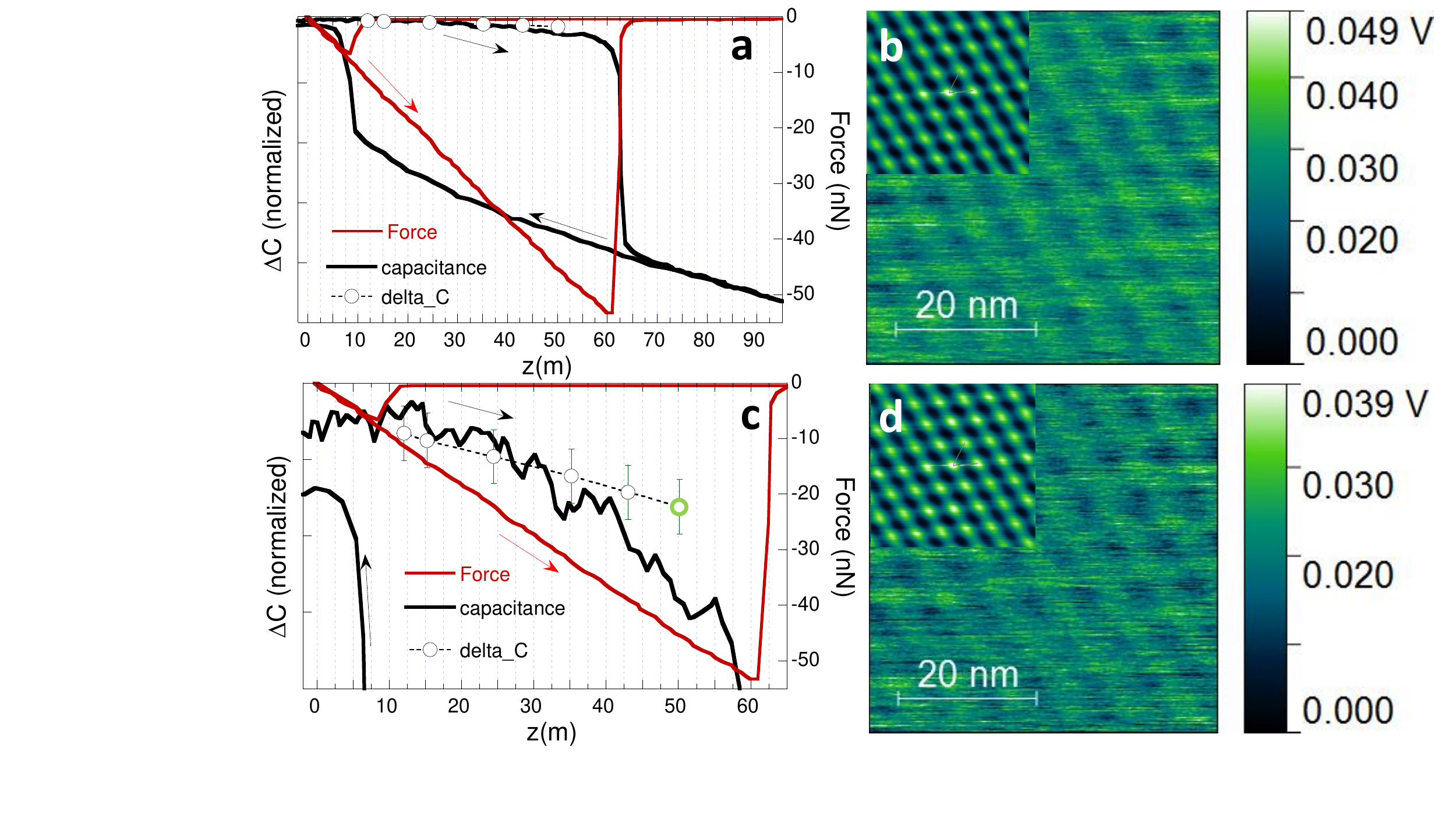}
  \caption{{\textbf NAP experiments: close up on figure 2:} \textbf{a} and \textbf{c} Force-curve (red line) and capacitance (black line) approach of a tip towards a TBG:hBN:Glass system. The empty circles correspond to capacitance changes evaluated from the difference in capacitance images taken in lift mode from the contact mode as a function of $\Delta_z$; the error bars correspond to the RMS values at each image; \textbf{b} conductance image in contact mode, \textbf{d} conductance image with a tip lift of 50 nm. The insets show the autocorrelation functions for both images.}
  \label{figS5}
\end{figure}

Figures \ref{figS5} \textbf{a} and \textbf{c} are expanded portions of figure~2\textbf{a} and \textbf{b}, to help visualize the capacitance decrease (black solid lines) and concurrent force (red solid lines) measurements during the lift mode. The filled circles in figure~\ref{figS5}\textbf{c} represent the capacitance drops at different lift heights, evaluated from the sMIM capacitance images. The first observation is that although for small lifts the tip-retraction curve (solid back line) is still in agreement with the capacitance change observed in lift mode, it does not decrease as notably. The interpretation for this difference is related primarily with meniscus formation dynamics \cite{Szoszkiewicz2005s, Carpentier2015s}. During the lift scan, the tip stays at a constant height thus providing more time for meniscus nucleation. The expected impact on the capacitance difference is consequently a smaller decrease because of a higher volume meniscus, as observed. Examining the conductance channel images displayed in figures~\ref{figS5}\textbf{b}, contact, and \ref{figS5}\textbf{d}, we observe that imaging is still adequately performed for a lift of 50~nm (also see the green circle in figure~\ref{figS5}\textbf{c}). The inset shows the auto-correlation functions, showing minimal changes in correlation lengths and same periodicity. The color scale nevertheless shows a decrease in signal amplitude for the image acquired in the lift mode, consistent with a more decoupled system because of a stretched meniscus.

\end{document}